\documentclass[referee, bibyear]{aa} 

\usepackage{graphicx}
\usepackage{txfonts}
\begin{document}

   \title{The $1997$ Event in the Crab Pulsar in X-rays}

   \author{M. Vivekanand}

   \institute{No. $24$, NTI Layout $1$\textsuperscript{st} Stage, 
              $3$\textsuperscript{rd} Main, $1$\textsuperscript{st} 
              Cross, Nagasettyhalli, Bangalore $560094$, India. \\
              \email{viv.maddali@gmail.com}
             }

   \date{Received: 15 Jul 2015; Accepted: 30 Dec 2015, as Research Note in A\&A}

 
  \abstract
   { In October $1997$, radio pulses from the Crab Pulsar underwent abnormal delay. 
     This was reported by two radio observatories, both of which explained this 
     frequency dependent and time varying delay as being due to refractive effects 
     of ionized shells in the Crab Nebula. Both groups also noted that, curiously 
     and confusingly coincident with the frequency dependent delay, the Crab Pulsar 
     also underwent an unusual slowing down, which they believed to be unrelated to 
     the Crab Nebula and instead intrinsic to the Crab Pulsar, resulting in an
     additional delay that was frequency independent. However, it now appears that 
     one of the groups attributes the frequency independent delay also to refractive 
     effects.
   }
   { This work aims to verify whether at least a part of the frequency independent 
     delay is indeed due to intrinsic slowing down of the Crab Pulsar.
   }
   { Timing analysis of the Crab Pulsar's October $1997$ event has been done in 
     X-rays, which are not delayed by the refractive and diffractive effects that 
     affect radio waves; at X-rays only the intrinsic slowing down should contribute
     to any observed delay. Data mainly from the PCA instrument aboard the RXTE 
     satellite have been used, along with a small amount of data from the PDS 
     instrument aboard the BeppoSAX satellite.
   }
   { Analysis of the X-ray data, using the very accurate reference timing model 
     derived at radio frequencies, strongly supports the intrinsic slowing down 
     hypothesis. Analysis using the reference timing model derived self-consistently
     from the limited X-ray data, which is less accurate, is not completely unambiguous 
     regarding the above two hypotheses, but provides reasonable support for the 
     intrinsic slowing down hypothesis.
   }
   { A significant fraction of the frequency independent delay during the October 
     $1997$ event is indeed due to intrinsic slowing down of the Crab Pulsar.
   }


   \maketitle
%

\section{Introduction}

Radio pulses from pulsars undergo several kinds of transformation while traveling 
through the ionized interstellar medium, such as dispersion, pulse scatter broadening, 
scintillation, and delayed arrival of pulses; see \cite{Manchester1977} and 
\cite{Lyne2006} for pedagogical reviews; see also  \cite{Backer2000} and \cite{Lyne2001} 
and references therein for technical details. The Crab Pulsar in particular displays 
extreme forms of such activity from time to time (\cite{Smith2000}). However, the event 
of late October $1997$ in the Crab Pulsar stands out in that it was simultaneously 
accompanied by a dramatic slowing down of the Crab Pulsar (\cite{Backer2000, Smith2000, 
Lyne2001}); it slowed down by $\approx 1.2$ ms in  $\approx 10$ to $20$ days. In
comparison, the Crab Pulsar  wanders in delay by about $\approx 0.3$ ms over a month owing 
to timing noise (\cite{Backer2000}), and is delayed by about $\approx 0.0007$ ms over 
$20$ days owing to systematic loss of rotational energy. \cite{Backer2000} and \cite{Smith2000} 
note the curious and confusing coincidence of the dramatic slowing down (which is presumed 
to be due to factors internal to the Crab Pulsar) with the dispersive, diffractive, and 
refractive phenomena (which are presumed to be due to factors related to the  Crab Nebula).  
Both groups developed their respective models to explain the dispersive, diffractive, and 
refractive phenomena in terms of moving ionized screens within the Crab Nebula 
(\cite{Backer2000, Smith2000, Lyne2001}. 

 More recently, however, \cite{Smith2011} have  suggested that there was no intrinsic 
slowing down of the Crab Pulsar during the October $1997$ event; they attribute the 
entire non-dispersive, frequency independent delay of $1.2$ ms to ray paths in highly 
localized electron clouds in the Crab Nebula. A similar hint appears in the work 
of \cite{Wong2001}, in the last sentence of their section 3.4. A perspective of the 
problem is provided by Figure $2$(a) in \cite{Smith2011}, which shows the variation 
of frequency independent timing residuals of the Crab Pulsar during the October $1997$ 
event. The event being discussed here refers to the dramatic slowing down of the Crab Pulsar at around MJD 50750 (29 October $1997$), which manifests as increasing timing 
residuals. A glitch occurred at MJD $50812.59 \pm 0.01$ (\cite{Espinoza2011}), which 
is not believed to be connected to the October $1997$ event (\cite{Backer2000, Smith2011}).

This work attempts to verify, at least qualitatively, whether the Crab Pulsar  
intrinsically slowed down or not 
during late October $1997$ by analyzing X-ray data for that epoch. X-ray propagation is 
not affected by passage through ionized media, unlike radio waves; so X-ray timing of the 
Crab Pulsar should show only the unusual intrinsic slowing down, if any, and not other 
delays. On the other hand, X-ray data is very sparse (i.e., poorly sampled in time) 
compared to radio data. Only four X-ray observatories capable of timing the Crab Pulsar 
observed it during that epoch. Out of the four, ROSAT and ASCA did not yield useful data. 
BeppoSAX provided useful data for only one epoch. All other data came from the RXTE 
observatory; however, these observations occurred on  average once  a week or once every 
two weeks, whereas the radio observations took place at least once every day. Therefore, 
quantitative comparison between the X-ray and radio results may not be constraining.


\section{Observations}

\subsection{RXTE PCA observations}

The RXTE data are obtained from the Proportional Counter Array (PCA; \cite{Jahoda1996}) 
aboard the RXTE observatory. The PCA consists of five proportional counter units (PCUs) 
operating in the $2$–$60$ keV range, having a field of view of $1^\circ$ in the sky, and 
a time resolution of $1$ $\mu$sec (see ``The ABC of XTE'' guide on the RXTE 
website\footnote{heasarc.gsfc.nasa.gov/docs/xte/data\_analysis.html}). The first 
observation was on $20$ July $1997$ (MJD $50650$, ObsID 10200-01-12-00) and the last 
observation was on $11$ April $1998$ (MJD $50914$, ObsID 30133-01-08-00); there were a totoal of  
$28$ ObsIDs. The closest observation preceding this duration was on $9$ May $1997$ (MJD 
$50577$, ObsID 10200-01-13-00), which was not included in the analysis owing to the $\approx 
73$-day gap. The last observation for the analysis was chosen so as to sufficiently represent 
the glitch behavior.  These data were obtained in the \textit{Generic Single Bit} mode, 
having the configurations \textit{SB\_250us\_0\_13\_2s} and \textit{SB\_250us\_14\_249\_2s}. 
Both configurations accumulated photon events into time bins of size $244.14$ $\mu$sec; 
the former mode combined the first $14$ energy channels into a single channel, while the 
latter combined the next $236$ energy channels. Of the $28$ ObsIDs, $22$ had all five PCUs 
switched on, while 3 had four PCUs switched on, and 3 had only three PCUs switched 
on.

\subsection{BeppoSAX PDS observation}

The BeppoSAX observatory observed the Crab Pulsar on $8$ October $1997$ (MJD $50729$, 
ObsID $20795004$) using, among other narrow field instruments (NFIs), the Phoswich 
Detector System (PDS;   \cite{Frontera1997}). The PDS consists of four NaI(Tl)/CsI(Na) 
phoswich scintillation photon counters grouped into two pairs, which rock on 
and off the source in the sky, similar to the HEXTE instrument of RXTE. It covers the 
energy range $15$–$300$ keV and has a time resolution of $16$ $\mu$sec (see the ``Cookbook 
for BeppoSAX NFI Spectral Analysis'' guide on the BeppoSAX 
website\footnote{http://www.asdc.asi.it/bepposax/software/index.html}). This lone  
BeppoSAX observation was considered important since its epoch is close to the dramatic 
slowdown of the Crab Pulsar. BeppoSAX also observed the Crab Pulsar on $6$ April $1998$ 
(MJD $50909$), which falls in the glitch part of the data in Figure~\ref{fig1}; this 
data could not be analyzed owing to the lack of housekeeping files.

%

\section{Analysis}

\cite{Vivekanand2015} has described in comprehensive detail the timing analysis for 
the Crab Pulsar using the HEXTE instrument. The method of analysis differs from 
instrument to instrument mainly up to the stage of referring the photon arrival times 
to the solar system barycenter.

When combining data from different instruments, the keyword \textit{TIMEPIXR} is 
important. All data used in this work have \textit{TIMEPIXR} $ = 0$.  The pulse 
arrival epochs determined by the tool \textit{efold} have been corrected by the values 
$+122$ $\mu$sec and $+8$ $\mu$sec for the instruments PCA/SB and PDS, respectively, 
as per the prescription in the document ``RXTE Absolute Timing Accuracy'' on the RXTE 
website\footnote{http://heasarc.gsfc.nasa.gov/docs/xte/abc/time.html}.

\subsection{RXTE PCA analysis details}

Data for each of the two configurations were analyzed separately until the very end.
It was found that in $25$ of the $28$ ObsIDs, there was negligible difference in 
arrival times of the two configurations (less than a milliperiod), so their average 
was used for the final analysis. The same was done for two of the three remaining ObsIDs, 
in which the difference was $-477.9994$ periods, which is very close to an even number 
of periods. In the third ObsID, the difference was $-89.0000$ periods, which is an odd 
number of periods, so averaging the two arrival times will result in an error of half a 
period; so the arrival time of one of the configurations was used in this case. For
more details of analysis for this instrument see \cite{Vivekanand2015}.

\subsection{BeppoSAX PDS analysis details}

BeppoSAX PDS data was analyzed using FTOOLS version $5.0$ and SAXDAS version $2.3.1$.
The following data was downloaded for ObsID $20795004$: the cleaned and linearized 
event file, the GTI file,  and the housekeeping file. Using the housekeeping file, an 
additional system related GTI was created using the following selections: (a) Earth 
occultaion angle, EARTH\_ANG > $14^\circ$; (b) time since the peak of the last South 
Atlantic Anomaly passage, TIME\_SINCE\_SAA $> 30$ min or TIME\_SINCE\_SAA $< 0$ min; 
(c) star tracker configuration, ID STR\_CONF >= 4 and STR\_CONF <= 6; see the 
``Aspect Reconstruction'' section of ``BeppoSAX Cookbook'' on the BeppoSAX 
website\footnote{http://www.asdc.asi.it/bepposax/software/cookbook/attitude.html}.
In summary, GTI selection criteria are as per standard procedures, with additional
constraints on the three parameters mentioned above. After selecting photons based 
on this GTI, the light curve was obtained using the tool \textit{lcurve}; based on 
this light curve, the data beyond $70$ kilo seconds from the start of the data file were excluded 
from further analysis. Then photons were selected in the energy range $15$ to $220$
keV.  Finally the tool \textit{baryconv} was used to refer photon 
arrival times to the solar system barycenter. Further analysis was similar to that 
of the PCA instrument of RXTE.

%

\section{Results for the October 1997 event using the radio reference timing model}

\begin{figure}[t]
\centering
\includegraphics[width=11cm]{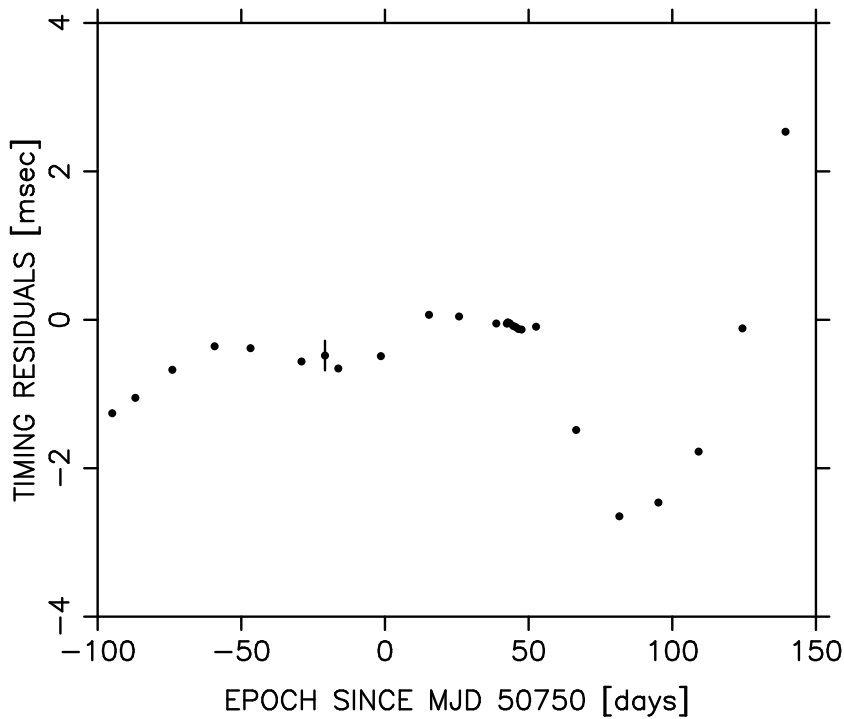}
\caption{Result of using TEMPO2 on all data with the parameters given in Table $1$ 
         of \cite{Backer2000} as constant input (i.e., without any fitting). The 
         lone BeppoSAX data point is at epoch $\approx -21$ days with a larger error bar.
         We note that no artificial phase jumps (using PHASE $\pm 1$ in the input file to 
         TEMPO2) are required, pointing to the robustness of the radio reference timing 
         model used.
        }
\label{fig1}
\end{figure}

In this section the X-ray data are analyzed using the reference timing model of 
\cite{Backer2000}, given in their Table $1$; this is used to check that   the same 
phenomenon is being observed that was observed by radio astronomers. The Jodrell group have not published 
their reference timing model. Figure~\ref{fig1} shows the result of using TEMPO2 along 
with the parameters in Table $1$ of \cite{Backer2000} as constant input (i.e., without 
any fitting) for the X-ray data from the PCA/HEXTE and from PDS/BeppoSAX. This figure 
is remarkably similar to Figure $2$(a) in \cite{Smith2011}. The slowing down of the 
$1997$ event and the glitch behavior are clearly evident, including the negative 
slope in residuals between the two epochs. In Figure~\ref{fig1}, the Crab Pulsar slowed 
down by $\approx 0.7 \pm 0.1$ ms over a duration of $\approx 30$ days around MJD $50750$. 
This is smaller than, but consistent with, the findings of \cite{Backer2000, Smith2000}, and \cite{Lyne2001}.  Furthermore, the delay observed here is about half of that observed at radio 
frequencies. We  therefore conclude  that the Crab Pulsar indeed 
slowed down intrinsically during the $1997$ event.

%

\section{Results for the October 1997 Event using the reference timing model derived from X-rays}

In this section the X-ray data are analyzed using reference timing models that are 
derived self-consistently  from the X-ray data itself. These data are not expected to be 
as accurate as the radio reference timing model owing to sparse X-ray timing data.

The first step is to identify precisely the ``post-event and pre-glitch'' duration in 
Figure~\ref{fig1}; this duration will be used to obtain the reference timing model for the 
entire data. Based on the independent identifications by 
\cite{Backer2000}, \cite{Lyne2001}, and \cite{Smith2011}, which differ marginally from 
each other, for our purposes the post-event and pre-glitch duration can be safely 
identified as being from MJD $50758$ to MJD $50812$.

The second step is to fit the X-ray data in the above post-event and pre-glitch duration 
to a pulsar slowdown model (using TEMPO2 \cite{Hobbs2006}), which is conventionally taken
as consisting of three parameters: rotation frequency $\nu$, its derivative $\dot \nu$, 
and its second derivative $\ddot \nu$; see Equation 1 of \cite{Espinoza2011}. Two such 
models are derived, which are shown as models $1$ and $2$  in Table~\ref{tbl1}. 
Model $1$ is obtained by a fit of the X-ray data in the post-event and pre-glitch 
duration using TEMPO2. Model $2$ is an arbitrary but minor modification of model $1$, for 
reasons explained later. The rms residuals of the fit in the two models is $21$ $\mu$sec 
and $68$ $\mu$sec, respectively. Within the errors both models are similar. The corresponding 
parameters from the Jodrell Bank monthly timing ephemeris of the Crab 
Pulsar\footnote{http://www.jb.man.ac.uk/pulsar/crab.html} (\cite{Lyne1993}), referring 
to the reference epoch in Table~\ref{tbl1}, are $29.870335055$ Hz and $-3.75274 \times 
10^{-10}$ s$^{-2}$, which compare very well with the estimates in Table~\ref{tbl1}.

\begin{table}
\caption{TEMPO2 best fit parameters to the post-event and pre-glitch duration in 
Figure~\ref{fig1}, which is identified as being from MJD $50758$ to MJD $50812$.
$\nu$ is the rotation frequency of the Crab Pulsar at the reference epoch MJD 
$50797.5233442733588$; $\dot \nu$ and $\ddot \nu$ are the first and second time 
derivatives of $\nu$, respectively, at the same epoch. The errors in brackets are 
in the last digit of each result. Both models use PCA data only, since the BeppoSAX
data lies outside this range. While model $1$ is obtained using TEMPO2, model $2$
is an arbitrary modification of model $1$, for reasons explained in the text, which
is why no errors are quoted for these parameters.
}
\label{tbl1}
\centering
\begin{tabular}{|l|c|c|}
\hline \hline
Parameter  & Model $1$ & Model $2$ \\
\hline
$\nu$ (Hz)  & $29.870335054(1)$  & $29.870335055$ \\
\hline
$\dot \nu$  ($10^{-10}$ s$^{-2}$) & $-3.75266(2)$ & $-3.75265$ \\
\hline
$\ddot \nu$  ($10^{-20}$ s$^{-3}$) & $0.8(2)$ & $1.2$ \\
\hline
\end{tabular}
\end{table}

\begin{figure}[b]
\centering
\includegraphics[width=11cm]{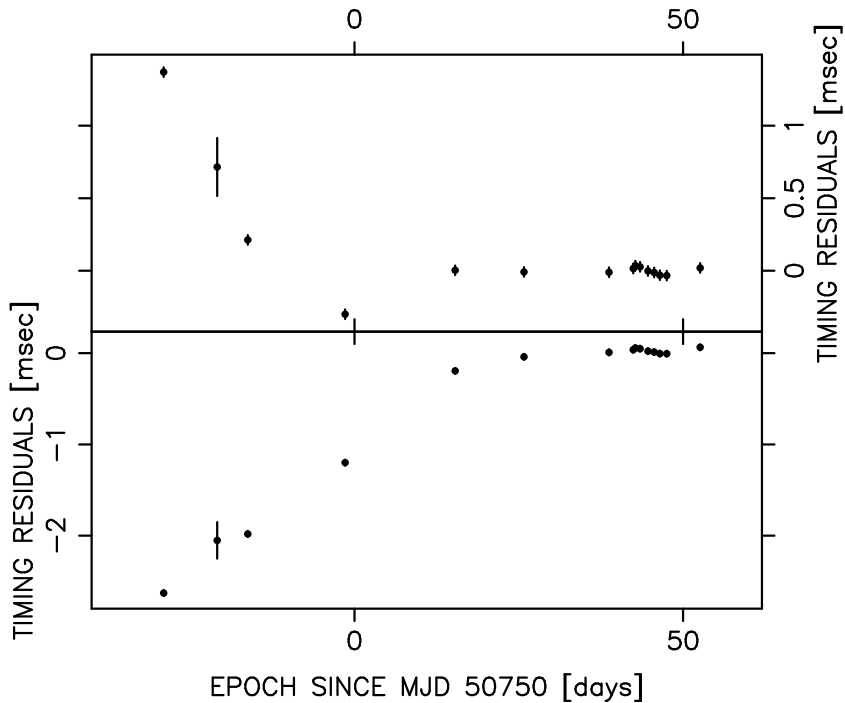}
\caption{Top panel: Result of using TEMPO2 on the X-ray data with the parameters of 
         model $1$ in Table~\ref{tbl1} as constant input (i.e., without any fitting).
         The plot is restricted to the post-event and pre-glitch duration, along 
         with four data points before this duration, to better view the event itself. Bottom 
         panel: Same as Top panel, but using the parameters of model $2$ in~\ref{tbl1}.
        }
\label{fig2}
\end{figure}

The final step is to use TEMPO2 along with the parameters in Table~\ref{tbl1} as 
constant input (i.e., without any fitting) for the entire data; the results are presented in 
the two panels of Figure~\ref{fig2}. The glitch behavior is clearly seen using both models,
although it is not shown in the figure; qualitatively they are similar to the glitch behavior 
seen in Figure $2$(a) in \cite{Smith2011}, while quantitative comparison is done in the 
following section. The occurrence of a glitch soon after the event of October 
$1997$ in the Crab Pulsar is fortuitous, since its analysis in the current X-ray 
data validates the method of analysis for the event as well.

Unfortunately, the X-ray data analysis of this section is not unequivocal regarding 
the October $1997$ event. Using Model $1$, the pre-event X-ray timing residuals can be 
interpreted as a speeding up of the Crab Pulsar (top panel of Figure~\ref{fig2}), while using Model $2$ they 
can be interpreted as a slowing down  (bottom panel of Figure~\ref{fig2}). 
The most probable reason for this is the sparse (i.e., poorly sampled in time) X-ray data. 
Model $2$ was obtained by trial and error, in the effort to ascertain whether the X-ray 
data could show an event behavior contrary to that obtained using Model $1$.  We 
therefore conclude from the above analysis that the X-ray data by itself cannot 
categorically determine whether the Crab Pulsar intrinsically slowed down during the event 
of October $1997$.

However, a closer look at Figure~\ref{fig2} reveals a detail. In both panels of
Figure~\ref{fig2}, the first data point before the post-event and pre-glitch 
duration at epoch $\approx -1.43$ days is lower than the reference data (i.e., 
data later than this epoch in Figure~\ref{fig2}). In the bottom panel of Figure~\ref{fig2}
it is expected to be so anyway, but in the top panel of Figure~\ref{fig2}  it 
is lower by $0.30 \pm 0.04$ ms. Therefore, it appears that even Model $1$ indicates a 
slowing down of the Crab Pulsar around the epoch $-1.43$ days, or equivalently at round 
MJD $50750 - 1.43 =$MJD $50748.57$. It can therefore be concluded from the above analysis that 
although the X-ray data analysis of this section cannot categorically determine whether 
the Crab Pulsar slowed down during the event of October $1997$, it offers reasonable 
support for that hypothesis.

%

In figures ~\ref{fig1} and ~\ref{fig2}, the lone BeppoSAX timing residual differs
by about $\approx 0.17 \pm 0.20$ ms with respect to RXTE/PCA data, which is
consistent with the formal error of $\approx 0.2$ ms on this timing residual
(\cite{Mineo2000}). See also \cite{Cusumano2003} and \cite{Nicastro2004} for 
details of the accuracy of the onboard clock of BeppoSAX. The lone BeppoSAX 
data point used in this work follows the RXTE timing residual variation in both 
figures ~\ref{fig1} and ~\ref{fig2}, without having to artificially insert phase 
cycle corrections in TEMPO2. We believe that the lone BeppoSAX data point 
significantly enhances the confidence of the fits to the RXTE data since it is 
obtained from an entirely independent observatory.

\section{Results for the December $1997$ glitch}

The results of the previous section depend critically upon the post-event and pre-glitch solutions of TEMPO2, which are listed in Table~\ref{tbl1}. An independent 
verification of these is to use them to derive the parameters of the glitch that 
soon followed the October 1997 event, and compare them with results obtained at 
radio frequencies. For this glitch, \cite{Wong2001} have publish more detailed parameters 
than \cite{Espinoza2011}; therefore, their method of glitch analysis has been followed. 
We  used their glitch epoch (MJD 50812.9) and their decay time scale ($2.9$ days) 
since the X-ray data is too sparse to derive these parameters independently. The X-ray 
data beyond the glitch epoch were  analyzed using the model of \cite{Backer2000}, 
and models $1$ and $2$ of Table~\ref{tbl1}.  The results are shown in Table~\ref{tbl2}; 
for comparison the corresponding values derived by \cite{Wong2001} are also shown 
(see their Table $3$). 

\begin{table}
\caption{TEMPO2 best fit parameters to the post-glitch X-ray data. The second column 
         lists the glitch parameters derived by \cite{Wong2001}. Their values of 
         glitch epoch $t_g$ and decay time scale $\tau_n$ have been used as given 
         constants. The derived 
         glitch parameters are (a) jump in frequency that is exponentially recovered 
         $\Delta \nu_n$, (b) permanent jump in frequency $\Delta \nu_p$, and (c) 
         permanent jump in frequency derivative $\Delta \dot \nu_p$. The third to 
         fifth columns list the values derived  using the model of \cite{Backer2000}, 
         and models $1$ and $2$ in Table~\ref{tbl1}.
}
\label{tbl2}
\centering
\begin{tabular}{|l|c|c|c|c|}
\hline \hline
Parameter  & \cite{Wong2001} & \cite{Backer2000} & Model $1$ & Model $2$\\
\hline
$t_g$ (MJD)  & $50812.9^{+0.3}_{-1.5}$ & &   &  \\
\hline
$\tau_n$ (days)  & $2.9 \pm 1.8$ & &   &  \\
\hline
$\Delta \nu_n$ ($10^{-7}$ Hz)  & $2.4 \pm 0.6$ & $2.1 \pm 0.9$ & $2.1 \pm 0.9$  & $2.1 \pm 0.9$ \\
\hline
$\Delta \nu_p$  ($10^{-7}$ Hz) & $0.17 \pm 0.05$ & $0.28 \pm 0.06$ & $0.32 \pm 0.06$ & $0.26 \pm 0.06$  \\
\hline
$\Delta \dot \nu_p$  ($10^{-15}$ s$^{-2}$) & $-14.2 \pm 0.6$ & $-13 \pm 2$ & $-8 \pm 3$ & $-14 \pm 2$   \\
\hline
\end{tabular}
\end{table}

The errors on $t_g$ and $\tau_n$ in column $2$ of Table~\ref{tbl2} cannot be used to 
estimate the errors in columns $3$ to $5$, because the parameters of the fit are highly 
correlated, and the covariance matrix is obtained only when $t_g$ and $\tau_n$ are also fit.
The actual uncertainties are larger than those quoted in columns $3$ to $5$ of table 2. However, 
the mean values of the derived X-ray parameters can be compared with the corresponding radio 
parameters and their errors.

From Table~\ref{tbl2} it is clear that the glitch parameters derived using the three
models are broadly consistent with each other. It can therefore be concluded 
that the post-event and pre-glitch solutions of Table~\ref{tbl1} are fairly robust, 
and enhance the confidence in the results of the previous section.

\section{Discussion}

We  conclude from this work that, strictly speaking, the X-ray data by itself cannot unequivocally discriminate 
between the two hypotheses, but offers reasonable evidence for intrinsic slowing down of the 
Crab Pulsar during the October $1997$ event. However, when coupled with the very accurate 
radio reference timing model, the X-ray data strongly supports the intrinsic slowing down 
hypothesis. It may be worthwhile to check the Jodrell data for other occurrences of such
slowing down in the Crab Pulsar.

Assuming that the Crab Pulsar indeed slowed down, the implications are the following:

\begin{enumerate}
\item The effect of such extreme events in pulsars with plerions must be taken into account 
while deriving their glitch parameters.
\item This is probably a slow version of the anti-glitch observed by \cite{Archibald2013} in
a magnetar.
\item The explanation of \cite{Akbal2015} for the observed slowing down of PSR J1119-6127 
after a glitch in May 2007 could be the mechanism driving the slowing down in the Crab Pulsar.
\end{enumerate}

\begin{acknowledgements}
I thank Francis Graham-Smith for useful and encouraging discussion, and Andrew Lyne for 
verifying the X-ray timing residual behavior. I thank the referee for useful comments. 
This research made use of data obtained from the High Energy Astrophysics Science Archive 
Research Center Online Service, provided by the NASA-Goddard Space Flight
Center. It has also made use of data provided by the BeppoSAX satellite.
\end{acknowledgements}


\end{document}